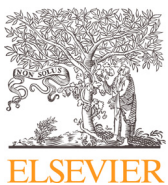
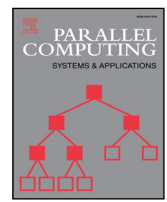

# iPregel: Vertex-centric programmability vs memory efficiency and performance, why choose?

Ludovic A.R. Capelli [a,*], Zhenjiang Hu [b], Timothy A.K. Zakian [c], Nick Brown [d], J. Mark Bull [d]

[a] *Institute for Computing Systems Architecture, University of Edinburgh, 10 Crichton Street, Edinburgh EH8 9AB, United Kingdom*
[b] *Information Systems Architecture Research, National Institute of Informatics, 2 Chome-1- Hitotsubashi, Chiyoda, Tokyo 100-0003, Japan*
[c] *Department of Computer Science, University of Oxford, 15 Parks Road, Oxford OX1 3QD, United Kingdom*
[d] *Edinburgh Parallel Computing Centre, University of Edinburgh, 47 Potterrow, Edinburgh EH8 9BT, United Kingdom*

**ABSTRACT**

The vertex-centric programming model, designed to improve the programmability in graph processing application writing, has attracted great attention over the years. Multiple shared memory frameworks that have implemented the vertex-centric interface all expose a common tradeoff: programmability against memory efficiency and performance.

Our approach consists in preserving vertex-centric programmability, while implementing optimisations missing from FemtoGraph, developing new ones and designing these so they are transparent to a user's application code, hence not impacting programmability. We therefore implemented our own shared memory vertex-centric framework iPregel, relying on in-memory storage and synchronous execution. In this paper, we evaluate it against FemtoGraph, whose characteristics are identical, but also an asynchronous counterpart GraphChi and the vertex-subset-centric framework Ligra. Our experiments include three of the most popular vertex-centric benchmark applications over 4 real-world publicly accessible graphs, which cover all orders of magnitude between a million to a billion edges. We then measure the execution time and the peak memory usage. Finally, we evaluate the programmability of each framework by comparing it against the original Pregel, Google's closed-source implementation that started the whole area of vertex-centric programming.

Experiments demonstrate that iPregel, like FemtoGraph, does not sacrifice vertex-centric programmability for additional performance and memory efficiency optimisations, which contrasts with GraphChi and Ligra. Sacrificing vertex-centric programmability allowed the latter to benefit from substantial performance and memory efficiency gains. However, experiments demonstrate that iPregel is up to 2300 times faster than FemtoGraph, as well as generating a memory footprint up to 100 times smaller. These results greatly change the situation; Ligra and GraphChi are up to 17,000 and 700 times faster than FemtoGraph but, when comparing against iPregel, this maximum speed-up drops to 10. Furthermore, on PageRank, it is iPregel that proves to be the fastest overall. When it comes to memory efficiency, the same observation applies; Ligra and GraphChi are 100 and 50 times lighter than FemtoGraph, but iPregel nullifies these benefits: it provides the same memory efficiency as Ligra and even proves to be 3 to 6 times lighter than GraphChi on average. In other words, iPregel demonstrates that preserving vertex-centric programmability is not incompatible with a competitive performance and memory efficiency.

## 1. Introduction

From social networks analysis to database queries, graph processing has become ubiquitous. The vertex-centric programming model introduced by Google in Pregel [1] offers a simple interface that enables users, such as data scientists, to easily develop graph processing applications. By exposing a set of highly-abstracted routines to the user, the vertex-centric interface greatly improves programmability. Through improved code readability and a less error prone code writing, the vertex-centric interface supports fast prototyping of graph applications. In the meantime, it requires no specialist programming expertise from the user, since the parallelisation and optimisations are offloaded to the underlying frame-work implementation. However, the vertex-centric programming model has traditionally been a compromise between the abstraction offered to the user and the performance achieved, as well as the memory footprint generated. To improve performance, certain vertex-centric frameworks abandon some features of the vertex-centric model. Whilst this improves performance, it can significantly impact the programmability and ease of use.

* Corresponding author.
*E-mail addresses:* l.capelli@ed.ac.uk (L.A.R. Capelli), hu@nii.ac.jp (Z. Hu), timothy.zakian@cs.ox.ac.uk (T.A.K. Zakian), n.brown@epcc.ed.ac.uk (N. Brown), m.bull@epcc.ed.ac.uk (J.M. Bull).

In order to support the processing of current and next generation graphs, vertex-centric models must maximise performance, so a key challenge is how to achieve this in a way that does not adversely impact the benefits of the vertex-centric model. The difficulty arises because the framework handles many aspects of the computation, from vertex selection to thread-safety, through inter-vertex communications.

*FemtoGraph*, for instance, is a framework that preserves the programmability of the vertex-centric interface. However, it does not efficiently handle the high volume of messages exchanged between vertices, which results in a large memory footprint. In addition, the performance observed is orders of magnitude worse than that of vertex-centric frameworks that make design choices in favour of performance by sacrificing programmability. There are certain frameworks, such as *GraphChi*, that do provide a better performance and memory efficiency by no longer abstracting certain computational steps from the user. *Ligra* provides an even better performance and memory efficiency, at the expense of an even greater programmability loss. Indeed, the programming model is no longer vertex-centric but vertex-subset-centric, where the user is required, for instance, to explicitly use parallelism and atomic operations.

By contrast, our approach focuses on improving performance and memory efficiency at no programmability cost. A major goal has been to design our framework to be optimisable without requiring application code rewriting for the user.

Our vertex-centric framework, *iPregel* [2], uses shared memory parallelism and in-memory storage. In this paper, we describe the novel features of *iPregel* in more detail and evaluate *iPregel* against vertex-centric[1] frameworks specifically designed for single-node execution too. We selected *FemtoGraph*, *GraphChi* and *Ligra* because they rely on fundamentally different designs. In this work, we do not only consider performance, but also the memory efficiency and general programmability. The results collected demonstrate that *iPregel* and *FemtoGraph* provide equally the best programmability. However, *iPregel* proves to be up to both 2300 times faster and 100 times more memory efficient than *FemtoGraph*. When compared against the frameworks optimised for performance, we observe that the memory footprint of *iPregel* is as small as that of *Ligra*, and up to 7.5 times smaller than *GraphChi*'s. In comparison to *GraphChi* and *Ligra*, the performance observed on *iPregel* varies for different benchmarks; for the Connected Components and SSSP benchmarks, *iPregel* is several times slower than *Ligra* and *GraphChi*, while remaining in the same order of magnitude. However, in PageRank, *iPregel* consistently exhibits the best performance above 4 threads, regardless of the graph. We thus demonstrate successful preservation of the vertex-centric programmability with no consequence on memory footprint. The impact on performance is greatly minimised compared to *FemtoGraph*, and for some benchmarks, *iPregel* even exhibits a greater performance than all frameworks tested.

The contributions described in this paper can be summarised as follows:

- **Programmability independent optimisations**, allowing the user to focus on application logic, then leverage the potential of optimisations without requiring code rewritings.
- **A lightweight and efficient** implementation of our framework *iPregel*, as well as a set of benchmark applications.
- **A thorough exploration** of vertex centric frameworks, including our own *iPregel*, contrasting three major attributes; performance, memory efficiency and programmability.

The rest of the paper is organised as follows: Section 2 presents related work and Section 3 briefly introduces the terminology used throughout this paper. Section 4 provides an overview of the *iPregel* framework, from its interface and implementation to the optimisations designed and how they can be leveraged by the user. Section 5 describes the other frameworks evaluated, followed by Section 6 which presents the application benchmarks selected. The conditions under which these experiments were run are outlined in Section 7. Finally, the results collected are presented, discussed and analysed in Section 8, before we draw conclusions in Section 9 and discuss potential further work directions.

## 2. Related work

First developed in 2010 through Pregel [1], the vertex-centric programming model has proven to be an intuitive way of developing graph processing algorithms. Many implementations have emerged, the majority of which rely on distributed memory architectures. Vertex-centric applications represent a real challenge to such architectures: they exhibit a high, and irregular, volume of communications as well as frequent global synchronisations. Nonetheless, with the exception of a few state-of-the-art machines that provide up to 160 terabytes of RAM[2] in a single memory space, only distributed memory architectures are able to provide enough RAM to process in-memory industry graphs, such as those employed by Facebook [3], which can contain up to a trillion edges.

To address this limitation, *GraphChi* [4] was developed; which is a single-node vertex-centric framework able to process graphs of any size. It does this via out-of-core computations; *GraphChi* relies on disk storage as an extension of memory. In this approach, the graph is divided into disjoint intervals, each of which is represented by a shard that stores all incoming edges of the vertices in that interval, on disk. *GraphChi* then loads shards[3] in turn into memory and processes concurrently the vertices belonging to the interval represented by the shard in memory. With this design, *GraphChi* is able to process graphs that do not fit in-memory, overcoming the fundamental limitation of single-node frameworks, at the expense of costly disk accesses.

*Ligra* [5], a shared memory framework using in-memory storage, has been a game changer with regard to the viability of in-memory execution of large graphs. Its authors argue that the amount of memory available in high-end single nodes is sufficient to process graphs with hundreds of billions of edges. Although this is true for certain state-of-the-art machines, realistically, as of 2018, a cluster node commonly has between 64 and 512GB of memory. Despite not meeting the memory needs of a trillion-edge graph, these nodes do provide enough memory to process graphs up to a hundred billion edges in *Ligra*. Therefore, single-node frameworks that rely exclusively on in-memory storage like *Ligra* can be viable also at large scale. *Ligra* demonstrates [5] that it is able to scale to graphs containing almost 13 billion directed edges, while preserving a parallel efficiency between 45% and 80% on 40 cores. From a programming perspective, *Ligra* is described as a vertex-subset-centric framework in [6]. This categorisation is motivated by the observation that *Ligra* retains a centralised view of

---

[1] And vertex-subset-centric.

[2] Record currently held by "The Machine", made by Hewlett Packard Enterprise.
[3] When determining the size of a shard, *GraphChi* ensures that it is sufficiently small to fit in-memory.

```
void IP_compute(struct IP_vertex_t* me);
void IP_combine(IP_MESSAGE_TYPE* old,
                IP_MESSAGE_TYPE new);
```

**Fig. 1.** User-defined functions of *iPregel*.

```
size_t IP_get_superstep();
bool IP_get_next_message(struct IP_vertex_t* me,
                         IP_MESSAGE_TYPE* msg);
void IP_send_message(IP_VERTEX_ID_TYPE destID,
                     IP_MESSAGE_TYPE msg);
void IP_broadcast(struct IP_vertex_t* me,
                  IP_MESSAGE_TYPE msg);
void IP_vote_to_halt(struct IP_vertex_t* me);
```

**Fig. 2.** Main functions (not all) provided by iPregel.

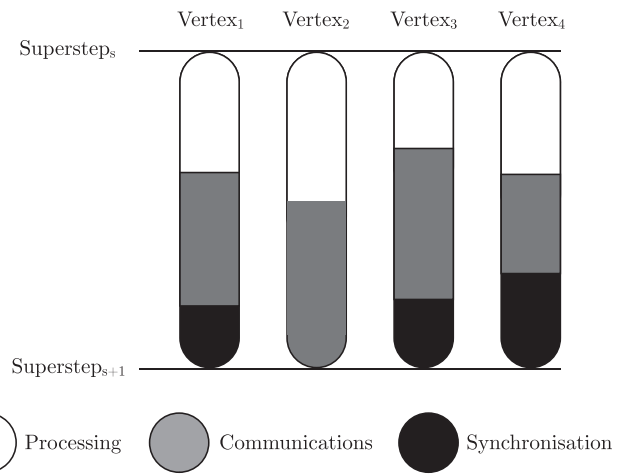

**Fig. 3.** A Bulk-Synchronous Parallel superstep.

the graph being processed, contrasting with a pure vertex-centric framework.

*FemtoGraph* [7] exposes the same characteristics as *Ligra*, namely shared memory parallelism and in-memory storage. The major difference resides in the fact that *FemtoGraph* preserves the vertex-centric programming model, exposed via its programming interface, which is identical to that of the initial Pregel. However, based on the results reported in [7], at best *FemtoGraph* provides little to no performance gain compared to existing graph processing frameworks. In addition, *FemtoGraph* appears to suffer from significant performance overhead at low number of threads.

## 3. Terminology

This section provides the reader with the technical terms that will be used throughout this paper.

A graph is made of vertices linked via edges that can be directed or undirected. Vertices standing at both ends of an edge are called *neighbours*. A directed edge linking the source vertex $v_{src}$ to the destination vertex $v_{dst}$ is said to be an *outgoing edge* of $v_{src}$ and an *incoming edge* of $v_{dst}$, abbreviated *out-edge* and *in-edge* respectively. Similarly, $v_{src}$ is known as an *incoming neighbour* of $v_{dst}$, abbreviated *in-neighbour*, and $v_{dst}$ is said to be an *outgoing neighbour* of $v_{src}$, or *out-neighbour*. *Adjacency list* is the term used to refer to the list of all neighbours of a given vertex. In directed graphs, it can be an *out-adjacency* list or *in-adjacency* list.

Finally, this paper will refer to the concept of *broadcast*, which in *iPregel* and other vertex-centric frameworks means sending a message to all out-neighbours of the broadcasting vertex.

## 4. Overview of iPregel

### 4.1. Interface

In *iPregel*, the user is provided with a simple Application Programming Interface (API), in which they must define the *combine* and *compute* functions, the signatures of which are illustrated in Fig. 1. The *compute* function contains the computation to execute on each vertex. The *combine* function is called when a vertex that has a mailbox already containing a message receives another message.

Fig. 2 illustrates supporting functions provided by *iPregel* that allow the user to track the superstep progression, read the messages received from last superstep, send a message to a specific out-neighbour or all out-neighbours at once, and halt the vertex currently processed respectively.

Although *iPregel* relies on shared memory parallelism, communications are achieved via a message-passing abstraction. Typically, message-passing abstractions are provided when writing codes for distributed memory architectures, nonetheless the motivation behind this choice for shared memory is multifaceted. Firstly, it protects the user from potential data races that arbitrary memory accesses could allow. Secondly, direct memory accesses require the programmer to know exactly where to write information, which implies exposing implementation details to the user. Finally, this abstraction provides *iPregel* with the freedom to optimise the underlying communication mechanisms whilst preserving a consistent interface to the user.

### 4.2. Implementation

The Bulk-Synchronous Parallel model (BSP [8]), on which *iPregel* relies, is illustrated in Fig. 3. This is a very common approach in vertex-centric processing, where the execution flow progresses in iterations, called supersteps, each made of three steps:

1. Local computation
2. Communications
3. Global synchronisation

In the context of vertex-centric programming, the first step consists in executing the user-defined function *compute* on each active vertex. During this phase, vertices can modify their state and have access to the messages they received during the previous superstep. During the second phase, the communications in Fig. 3, vertices send messages to their out-neighbours. Finally, once every active vertex has been processed and every outgoing message delivered, the superstep completes and a new superstep begins (synchronisation in Fig. 3).

As mentioned, *iPregel* applies the *compute* function defined by the user to each active vertex, at every superstep. When a vertex receives more than one message, the user's *combine* function is applied to combine messages on-the-fly. Under the hood, the message-passing abstraction provided to the user is implemented as direct memory accesses by *iPregel*. We argue that this gives the best of both worlds, it allows the user to rely on a simple interface for communication while exploiting the shared memory performance of a single-node solution.

From a parallelisation perspective, the *iPregel* framework is developed in C and relies on OpenMP [9] to support shared memory concurrency. Vertices are stored in a global array, and the list of their neighbour identifiers (known as their adjacency list) in another global array, both of which are shared by all threads. The vertex workload is then distributed using the default static schedule in OpenMP, in other words, the total number of vertices is evenly distributed across all threads and no work-stealing strategy is used.

Section 4.3 presents the optimisations strategies employed by *iPregel*. They all abide to the overarching philosophy of *iPregel*: optimisations should not require user source code rewriting or make the code more error prone, and implementation details related to the optimisations must stay abstracted from the user. This motivated the design of *iPregel* to trigger optimisations via compilation flags, which leaves the user source code untouched.

### 4.3. Optimisations

The optimisations presented in this section are not mutually exclusive. In other words, *iPregel* handles any combination of optimisations, without requiring additional work from the user, only a change of compilation flags is needed.

#### 4.3.1. Selection bypass

The first phase in vertex-centric frameworks consists in selecting the vertices to execute: this is already known to be a tricky aspect of vertex-centric models [10]. The naive approach is to check the status of each vertex and process those that are active.

However, for inactive vertices these checks are unfruitful and result in wasted memory accesses. This is important because frameworks that use in-memory storage and shared memory parallelism already place a high pressure on memory bandwidth. Therefore, keeping unproductive memory accesses to a minimum prevents aggravating that pressure. The naive approach becomes especially problematic in programs that contain a small number of active vertices, resulting in many wasted checks.

Thus we analysed the selection phase, which typically decides to run a vertex if it at least one of the following conditions is met:

1. It is the first superstep
2. The vertex is already active (that is, it did not halt when it was last processed)
3. The vertex received a message during previous superstep

Condition 1 becomes false at the end of the first superstep. Thus, from the second superstep onwards, a vertex is active if and only if conditions 2 or 3 are met. One cannot assert which condition it is, unless the algorithm exhibits what we refer to as *systematic halt*: every time a vertex is processed it halts at the end of the compute function. In other words, this algorithmic particularity guarantees that condition 2 is always false. It is the case in the Connected Components and SSSP benchmarks presented in Sections 6.2 and 6.3 respectively. By contrast, in the PageRank benchmark presented in Section 6.1, a vertex processed will not halt if the number of supersteps elapsed is less than the predefined threshold. In other words, in PageRank a vertex may be active in superstep $n$ without having received a message in superstep $n - 1$.

In the *systematic halt* situation however, this configuration is not possible. Indeed, since condition 1 is false after the first superstep and condition 2 is always false, only condition 3 remains; a vertex is active if, and only if, it received a message in previous superstep (as depicted in Fig. 4). Thus, the list of active vertices for superstep $n + 1$ can be established by monitoring message exchanges during superstep $n$ and finding which vertices are the recipients of these exchanged messages. This is why when an algorithm exposes *systematic halt*, *iPregel* can monitor message exchanges and automatically determine which vertices to run next superstep.

Integrating the *systematic halt* feature is straightforward; it can be embedded in the function called by vertices to send messages (*ip_send_message* and *ip_broadcast* as given in Fig. 2). The modification consists in appending the recipient vertex identifier to the list of vertices to run during next superstep. However, one must avoid duplicate identifiers so that a given vertex is not processed multiple times. This is again straightforward because when a vertex sends a message, the thread that runs that vertex must check if the recipient vertex already has a message in its mailbox, to determine whether it should apply the message combination presented in Section 4.3.3. From there, integrating the *systematic halt* feature consists of the thread adding the recipient vertex identifier to the list of vertices to execute during next superstep if that recipient vertex's mailbox is empty. Multiple threads accessing the same list can raise data races, this is why in *iPregel*, each thread maintains its own list. At the end of every superstep, these lists are merged into a single one. Then, one only needs to process the vertices in that list, without having to check their active status or the presence of pending messages. To exploit parallelism, this list is split evenly across all threads.

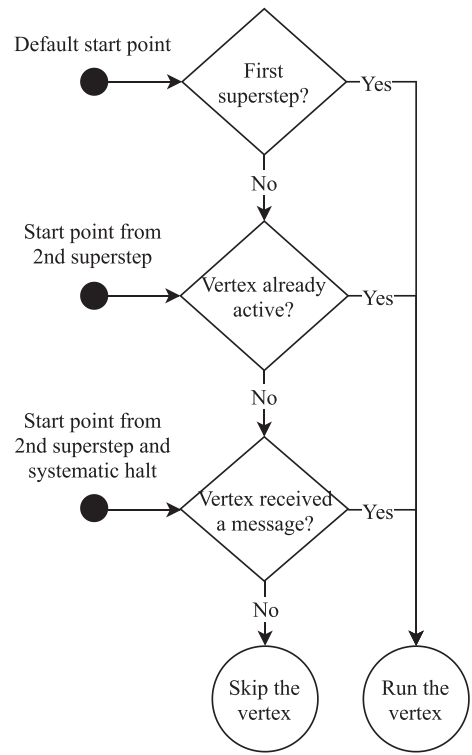

**Fig. 4.** Execution flow of the vertex selection mechanism.

There are multiple benefits from our selection bypass technique. Firstly, not having to check each individual vertex saves memory accesses and removes possible branch mispredictions on the vertex active state (execute if active, skip if inactive) since vertices in the merged list are known to be active. Secondly, this feature improves load balancing because threads receive exclusively vertices that are guaranteed to be run. This contrasts with the naive approach, where threads may receive identical numbers of vertices, potentially containing drastically different proportions of inactive vertices. In other words, our technique of selection bypass makes the active vertex distribution optimal with regard to the number of active vertices per thread.

#### 4.3.2. Message exchange

Typically, vertex-centric applications are communication intensive. Optimising the message exchange mechanism can therefore result in substantial improvements in performance.

There are two means by which a message can be transmitted: the sender can push it to a recipient mailbox, or the recipient can pull it from a sender's outbox. The push version can result in race conditions in the event of multiple vertices pushing to the same recipient mailbox concurrently. In *iPregel*, this is prevented with the use of busy-waiting locks. These are more efficient than their block-waiting counterparts, given that the combination operation is typically very small. However, the push version can be implemented without locks at all if the combination operation conveniently corresponds to an atomic operation. In *Ligra* [5], the user can exploit lock-free combination by writing a second, atomic, implementation of their combiner. Providing this optimisation without involving additional code writing may require a code parsing phase from the framework to determine whether a given combiner code can be atomically processed. In its current state, *iPregel* does not leverage the lock-free combiner optimisation.

The pull-based approach, due to the read-only nature of potential inter-thread interactions, has the advantage of being data-race free. Thus, threads can process message exchanges in parallel with no synchronisation. Vertices must have a mailbox to receive messages, as well as an outbox in which they can buffer the messages to send. Each message in the outbox must be attached with the recipient identifier, so that each out-neighbour knows which message take, if any. Such an approach would result in a heavy memory overhead, unless the same value is sent to all out-neighbours via a broadcast.

In that case, only one message needs to be stored in the vertex outbox, with no need to store the recipient identifier, since every out-neighbour is meant to read that message. It was observed that this optimisation can be applied to the majority of vertex-centric applications, since communications are typically performed via broadcasts to neighbours. This optimisation assumes that at most one broadcast is issued per vertex per superstep.

However, in order to support this lock-free design, *iPregel* must check, for each vertex, the outbox of every out-neighbour, which results in numerous memory accesses. In applications that expose a low number of active vertices, this optimisation generates a high number of memory accesses that consist in checking an empty outbox, hence wasting memory bandwidth and generating unproductive extra work. Although *Ligra* can dynamically switch between the push and pull communications at runtime via a threshold defined by the user, *iPregel* must be told whether to use the former or the latter via a compilation flag. The user must therefore determine experimentally whether it is beneficial in their case to enable this optimisation.

#### 4.3.3. Message combination

Vertex-centric frameworks that complete communications via message-passing provide each vertex with a mailbox. Messages received are then queued in the recipient mailbox. Due to the high-volume of messages exchanged in vertex-centric applications, this design eventually results in large mailboxes that no longer fit in-memory. This is where the concept of a combiner can be leveraged.

In the vast majority of vertex-centric applications, the user is interested in the sum, average, minimum or maximum value of the messages received. These happen to be both associative and commutative operations. Therefore, rather than queuing messages and then combining them in the user's *compute* function, the user can declare a combination operation that will be applied by the framework on-the-fly as messages arrive in the mailbox. The order in which messages are combined does not have to be enforced due to the commutativity property of the combination operation. The associativity allows for parallel reductions in a multi-threaded and/or distributed environment.

Combiners result in a significant reduction of mailbox size because each mailbox now stores a single message. Moreover, the framework no longer needs to dynamically resize mailboxes to fit new incoming messages. This saves numerous reallocations that can potentially become costly in a multi-billion edge graph. Furthermore, combiners expose additional optimisation opportunities to the framework, which can now optimise and parallelise the combination process.

```
void ip_combine(IP_MESSAGE_TYPE* old,
                IP_MESSAGE_TYPE  new) {
  if(*old > new)
    *old = new;
}
```

**Fig. 5.** The implementation in iPregel of a combiner keeping only the minimum value of messages received.

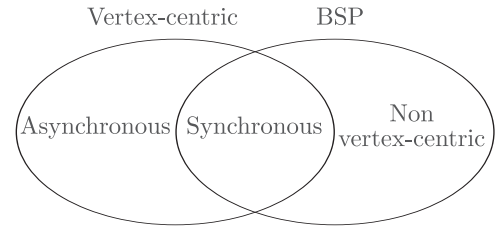

**Fig. 6.** Comparison of vertex-centric and BSP models of computation.

To use this feature, the *iPregel* user must define their own *combine* function and implement the operation to apply every time a new message is received. An example of a combiner calculating the minimum value of messages received is given in Fig. 5.

### 5. Frameworks considered

To evaluate *iPregel*, we consider in this paper only frameworks that are designed specifically for graph processing in shared memory. Nevertheless, Fig. 6, which is taken from McCune et al. [6], highlights three types of frameworks that could be considered. Among these categories, *iPregel* belongs to the middle; combining vertex-centric programming and synchronous execution. In order to provide a comparison of *iPregel* against a variety of frameworks, one framework of each type has been selected.

#### 5.1. GraphChi

The first vertex-centric framework to offer out-of-core computations [4], *GraphChi* belongs to the category of vertex-centric frameworks that exploit asynchronous execution, shown on the left in Fig. 6. For a graph that can fit entirely in memory, the out-of-core nature of *GraphChi* makes comparisons with an in-memory framework, such as *iPregel*, unfair. Fortunately, it turns out that *GraphChi* provides in-memory implementations[4] of its algorithms, which are automatically chosen by *GraphChi* when memory allows. Concretely, when running a *GraphChi* application, the user can pass the amount of RAM available via a runtime parameter. *GraphChi* then estimates the amount of memory needed for its in-memory version, and selects it if the memory available is sufficient. Experiments presented in this paper use the in-memory version of the implementations provided by *GraphChi*, the exception being the Single Source Shortest Path (SSSP) for which no implementation at all is provided. We therefore developed an in-memory implementation for SSSP, given in Fig. 7.

Another particularity of this framework is its asynchronous execution flow, where vertex updates are immediately visible, unlike its synchronous counterparts where updates take effect only in the following superstep. The advantage of the former is to help reach convergence faster, while the latter is easier to reason about by providing clearer semantics.

---

[4] Available at https://github.com/GraphChi/graphchi-cpp.

```
void update(graphchi_vertex<VertexDataType,
  EdgeDataType> &v, graphchi_context &ginfo) {
  assert(ginfo.scheduler != NULL);
  if(ginfo.iteration == 0) {
    set_data(v, vertex_values[v.id()]);
    if(v.id() == SOURCE_VERTEX)
      for(int j = 0; j < v.num_outedges(); j++)
        ginfo.scheduler->add_task(
          v.outedge(j)->vertex_id());
  } else {
    vid_t curmin = v.get_data();
    for(int i = 0; i < v.num_inedges(); i++)
      if(curmin > neighbor_value(v.inedge(i)))
        curmin = neighbor_value(v.inedge(i));
    if(curmin < v.get_data() - 1) {
      curmin++;
      set_data(v, curmin);
      for(int i = 0; i < v.num_outedges(); i++)
        if(curmin<neighbor_value(v.outedge(i))-1)
          ginfo.scheduler->add_task(
            v.outedge(i)->vertex_id());
    }
  }
}
```

**Fig. 7.** Compute function for SSSP in GraphChi.

```
void ip_compute(struct ip_vertex_t* me) {
  if(ip_is_first_superstep()) {
    me->value = initial_value;
  } else {
    IP_MESSAGE_TYPE sum = 0.0;
    IP_MESSAGE_TYPE msg;
    while(ip_get_next_message(me, &msg))
      sum += msg;
    msg = ratio + 0.85 * sum;
    me->value = msg;
  }

  if(ip_get_superstep() < 10)
    ip_broadcast(me, me->value
                     / me->out_neighbour_count);
  else
    ip_vote_to_halt(me);
}

void ip_combine(IP_MESSAGE_TYPE* old,
                IP_MESSAGE_TYPE  new) {
  *old += new;
}
```

**Fig. 8.** PageRank implemented in iPregel.

### 5.2. FemtoGraph

*FemtoGraph*[7] is a shared memory vertex-centric framework that uses exclusively in-memory storage and synchronous execution. It thus belongs to the middle category shown in Fig. 6, like *iPregel*. However, *FemtoGraph*[5] is designed and hard-coded for PageRank, in which all vertices are run at every superstep. As a consequence, *FemtoGraph* does not implement a vertex selection mechanism: it runs each vertex at every superstep, without checking their active status or the presence of pending messages in its mailbox. By contrast, the Connected Components and SSSP benchmarks do require vertices to be selected since they may become inactive during the computation. As a result, such algorithms cannot be implemented in *FemtoGraph* without rewriting parts of the framework itself. Also, we have not been able to observe correct results across all the graphs tested. Nonetheless, *FemtoGraph* remains an interesting reference since it is the only other vertex-centric framework specifically designed for in-memory storage and synchronous execution, like *iPregel*.

### 5.3. Ligra

Out of the three categories illustrated in Fig. 6, *Ligra*[6] belongs to the rightmost: non-vertex-centric frameworks, which includes vertex-subset-centric, with synchronous execution. Its approach, described as vertex-subset-centric in [6] as opposed to vertex-centric, consists in dividing the graph processed into subsets, which are run in turn. *Ligra* executes on each subset two functions defined by the user: one to apply to every vertex and one to apply to every edge. In addition, the user must implement the *compute* function, which in *Ligra* is the function that defines the overall execution flow of the application, from a graph-centric view. For instance, the user is in charge of writing the main loop, as well as its termination condition, within which they must explicitly pass the graph to the vertex and edge functions they defined earlier. Nonetheless, *Ligra* is a graph processing framework that relies on shared memory parallelism, in-memory storage and synchronous execution, so in that regard it acts as the non-vertex-centric counterpart of *iPregel*.

---

[5] Available at https://github.com/DataSys-IIT/FemtoGraph.
[6] Available at https://github.com/jshun/ligra.

## 6. Benchmark applications

In this work, we evaluate frameworks across three applications, namely PageRank, Connected Components and the Single-Source Shortest Paths. These three applications are widely used in vertex-centric experiments and thus act as standard benchmarks.

### 6.1. PageRank

Initially presented in [11], the PageRank algorithm is designed to order web pages based on their importance calculated from the number of hyperlinks pointing to them.

The *iPregel* implementation of a PageRank algorithm presented in Fig. 8 is based on the original Pregel version introduced in [1]. During the first superstep, each vertex begins with an initial PageRank value of one divided by the number of vertices, and broadcasts (as defined in Section 3) its PageRank value divided by its number of out-neighbours. From the next superstep onwards, each vertex sums the PageRank values received from its in-neighbours, then it updates its current PageRank value and broadcasts it again as described earlier. This is repeated for a pre-defined number of supersteps, after which vertices halt and the execution terminates. In practice however, a PageRank application would typically run until convergence is reached.

As explained above, when a vertex receives messages, it sums their values; this is an operation both associative and commutative, and so a combiner can be used as explained in Section 4.3.3. In Fig. 8, the reader can see that implementing this combiner requires very little work from the user: defining the *combine* function and writing a single line of code representing the sum to apply. In addition, the communications performed during the PageRank calculations consist exclusively of broadcasts, with a maximum of one broadcast per vertex per superstep. According to Section 4.3.2, this characteristic makes PageRank compatible with the pull-based communication model, which the *iPregel* design allows the user to enable just by passing a compilation flag. However, since vertices

```
void ip_compute(struct ip_vertex_t* me) {
  if(ip_is_first_superstep()) {
    me->value = me->id;
    ip_broadcast(me, me->value);
  } else {
    IP_MESSAGE_TYPE old_value = me->value;
    IP_MESSAGE_TYPE msg;
    while(ip_get_next_message(me, &msg))
      me->value = min(me->value, msg);
    if(me->value < old_value)
      ip_broadcast(me, me->value);
  }
  ip_vote_to_halt(me);
}

void ip_combine(IP_MESSAGE_TYPE* old,
                IP_MESSAGE_TYPE  new) {
  if(*old > new) *old = new;
}
```

**Fig. 9.** Connected components implemented in iPregel.

```
void ip_compute(struct ip_vertex_t* me) {
  if(ip_is_first_superstep()) {
    if(me->id == SSSP_SOURCE) {
      me->value = 0;
      ip_broadcast(me, me->value + 1);
    }
    else
      me->value = INF;
  } else {
    IP_MESSAGE_TYPE mindist = INF;
    IP_MESSAGE_TYPE msg;
    while(ip_get_next_message(me, &msg))
      mindist = min(mindist, msg);
    if(mindist < me->value) {
      me->value = mindist;
      ip_broadcast(me, me->value + 1);
    }
  }
  ip_vote_to_halt(me);
}

void ip_combine(IP_MESSAGE_TYPE* old,
                IP_MESSAGE_TYPE  new) {
  if(*old > new) *old = new;
}
```

**Fig. 10.** Unweighted SSSP implemented in iPregel.

halt only after a certain number of supersteps, in contrast to halting at every superstep, PageRank is not compatible with the *iPregel* selection bypass optimisation presented in Section 4.3.1.

### 6.2. Connected components

Computing the Connected Components of a graph consists in finding all disjoint subsets of that graph such that each subset is made only of vertices that can all reach one another. There are several possible vertex-centric algorithms to compute the Connected Components. The algorithm selected in *iPregel* is often referred to as Hash-Min. It relies upon the propagation of vertex identifiers to find, for each vertex, the minimum vertex identifier reachable. This computation converges, and thus terminates, when vertices find the minimum vertex identifier they can reach. Finally, vertices having reached the same minimum vertex identifier belong to the same Connected Component.

The *iPregel* implementation, illustrated Fig. 9, begins with vertices initialising their value to their own vertex identifier, before broadcasting it to out-neighbours. From then, vertices find the minimum vertex identifier received from their in-neighbours. They then update their value if the minimum vertex identifier obtained is smaller, in which case they broadcast it back to their out-neighbours to continue the propagation. Since vertices may obtain the minimum vertex identifier reachable at any superstep, they always halt at the end of a superstep.

The broadcast characteristic makes this Connected Components implementation compatible with the pull-based communications explained in Section 4.3.2. Also, the combination applied to messages received is once again an operation that is associative and commutative: the minimum. Therefore, the use of combiners can be leveraged as discussed in Section 4.3.3. Similarly to the PageRank combiner implemented in Fig. 8, the Connected Components equivalent is once again a *combine* function made of a single line of code. Furthermore, the fact that vertices halt at the end of every superstep makes the Connected Components also suitable for the selection bypass optimisation presented in Section 4.3.1, which too is enabled just by passing a compilation flag.

### 6.3. Single-source shortest paths

Finding Shortest Paths in graphs has many applications, as explained in [1]. In this work, we consider the Single-Source version of the Shortest Paths, where a vertex is selected as the source and the algorithm finds the minimum distance between that source vertex and every other vertex in the graph. In this benchmark we assume edge weights equal to 1.

The *iPregel* implementation given in Fig. 10 is based on the original Pregel version introduced in [1], which is considered as a distributed version of the Bellman-Ford algorithm [6], and is also the implementation used in *Ligra* for instance. During the first superstep, the source vertex initialises its value to 0 and begins the propagation by broadcasting its value incremented by 1 (representing the edge weight assumed). In the meantime, other vertices initialise their value to INF (a value greater than the longest distance possible in the graph). From the second superstep onwards, vertices calculates the potential minimum distance obtained from messages received. In the event this distance is smaller than the current vertex value, the vertex updates its value and broadcasts it incremented by 1 (representing the edge weight assumed). Finally, vertices halt at the end of every superstep since they may obtain their final minimum distance at any superstep.

This SSSP algorithm exposes the same characteristics as the Connected Components; vertices halt at the end of every superstep, communications are performed only via broadcasts, with a maximum of one broadcast per vertex per superstep, and it contains a combination operation that is associative and commutative (calculating the minimum). As a consequence, the SSSP implementation can be optimised with the selection bypass technique presented in Section 4.3.1, the pull-based communications discussed in Section 4.3.2 and the leverage of combiners introduced in Section 4.3.3 respectively.

Note that we amended, when needed, the benchmark implementations in other frameworks to become algorithmically equivalent such as homogenising the pre-defined iteration number in PageRank for instance.

## 7. Experimental environment

Experiments are run on Cirrus, an HPE/SGI Apollo 6800 system, in which each compute node is equipped with two 18-core Intel Xeon E5-2695 (Broadwell) series processors. Each compute node also has 256GB of RAM made of two Non-Uniform Memory Ac-cess (NUMA) regions of 128GB. Instances are set-up with CentOS 7 Linux operating system.

**Table 1**
Graphs selected (Abbreviations used: |V| = number of vertices, |E| = number of edges.

| Name | |V| | |E| |
|---|---|---|
| DBLP | 317,080 | 1,049,866 |
| Live Journal | 4,036,538 | 34,681,189 |
| Orkut | 3,072,441 | 117,185,083 |
| Friendster | 65,608,366 | 1,806,067,135 |

**Table 2**
Minimum, average and maximum speed-up of Ligra over iPregel when processing the Connected Components of each graph, across all numbers of threads tested.

| Graph | Min | Avg | Max |
|---|---|---|---|
| DBLP | 5.47 | 8.07 | 10.44 |
| Live Journal | 7.52 | 8.17 | 9.43 |
| Orkut | 5.72 | 6.47 | 7.77 |
| Friendster | 4.60 | 4.99 | 5.92 |

*iPregel* is compiled with *gcc* version 6.3.0, using C99 standard (GNU99 extensions when using spinlocks) and is parallelised with OpenMP. *Ligra* supports OpenMP and Cilk Plus parallelisation. In order to make the comparison with *iPregel* as consistent as possible, the OpenMP version was selected. The frameworks *Ligra*, *GraphChi* and *FemtoGraph*, which are developed in C++, are compiled with *g++* version 6.3.0, using C++14 standard. The optimisation level is set to −O3 for all frameworks.

The timings reported include only the processing time, that is, graph loading and dumping are not included. The second factor we have measured during experiments is the *resident set size*, which represents the peak memory usage of an application over its entire runtime. Unlike the performance, which is assessed purely on the runtime, the memory peak usage includes all phases of an application: graph loading, processing and dumping. The motivation is to assess whether a framework can process a graph given a certain amount of memory, which is conditional upon the success of all phases, not only the processing.

Table 1 lists the graphs processed in the experiments conducted in this work, where |V| is the number of vertices, and |E| the number of edges. They are real-world graphs selected from the online collection Stanford Network Analysis Project [12] (SNAP) and cover all orders of magnitude from million-edge to billion-edge. These graphs are undirected, thus the total number of directed edges is twice the amount presented. The smallest graph, the Database and Logic Programming Bibliography graph[7] (DBLP), is a real-world graph that represents the eponymous computer science bibliography. LiveJournal[8], Orkut[9] and Friendster[10] are network graphs; about blogging, social and gaming respectively.

## 8. Results

### 8.1. Performance

Fig. 9 illustrates the results of the three benchmarks across the four different frameworks, with different graphs. For PageRank, illustrated in the left column of Fig. 11, we observe that the *iPregel* version is 70 to 2300 times faster than its *FemtoGraph* counterpart.[11] *GraphChi* and *Ligra* outperform *FemtoGraph* too, exhibiting a maximum speedup of 700 and 17,000 respectively. The best sequential performance, regardless of the graph, is achieved by *GraphChi*[12], due to its asynchronous execution that enables vertices to read values updated by other vertices during this same superstep. However, it offers no performance gain when the number of threads increases and eventually falls behind both *iPregel* and *Ligra* versions. The results reported for PageRank in Fig. 11 demonstrate that the thread scalability[13] of *iPregel* is similar to that of *Ligra*. They also point to better graph scalability[14] in *iPregel*. In addition, Fig. 11 shows that the bigger the graph, the better the thread scalability of *iPregel*. Thenceforth, the performance differences observed for PageRank between *Ligra* and *iPregel* can be explained using these three factors. On the smallest graph DBLP, *Ligra* begins with a sequential runtime lower than *iPregel* and also provides a better thread scalability. Moving to the Live Journal graph, the number of vertices and edges are multiplied by 10 and 30 respectively. We can see that *iPregel* now begins ahead of *Ligra* at 1 thread, and provides a thread scalability better than in DBLP. Despite this progress, the strong thread scalability of *Ligra* eventually allows it to outperform *iPregel* above 8 threads. When moving to the hundred-million-edge graph Orkut, however, with PageRank, *iPregel* outperforms once again *Ligra* at 1 thread and manages to remain ahead across all numbers of threads thanks to its thread scalability improving. When moving to the billion-edge graph Friendster, *iPregel* now provides a thread scalability as good as that of *Ligra*, which allows its runtime remain to half that of *Ligra* across all numbers of threads. Overall, for PageRank, the best performance at 32 threads is achieved by *iPregel*.

---

[7] https://snap.stanford.edu/data/com-DBLP.html.
[8] https://snap.stanford.edu/data/com-LiveJournal.html.
[9] https://snap.stanford.edu/data/com-Orkut.html.
[10] https://snap.stanford.edu/data/com-Friendster.html.
[11] *FemtoGraph*'s timings for Orkut and Friendster graphs could not be collected due to abnormal termination and out-of-memory failure respectively.
[12] *GraphChi*'s timings for the Friendster graph could not be collected due to the number of file descriptors needed, approximately 21,000, being beyond our allowed limit.

For the Connected Components, whose results are shown in the middle column of Fig. 11, we find certain patterns already observed for PageRank. Namely, *GraphChi* offers no thread scalability, which allows *Ligra* and *iPregel* to become competitive at high number of threads. However, the performance achieved by *GraphChi* thanks to its asynchronicity is almost never equalled by *iPregel*, even at 32 threads. Nonetheless, *iPregel* continues to exhibit a better graph scalability than *Ligra* as illustrated in Table 2, in which we can see that the speed-up of *Ligra* over *iPregel* decreases as the graph grows, albeit always remaining greater than 1. Still, the vertex-centric *iPregel* remains up to 10 times slower than the vertex-subset-centric *Ligra* which leverages atomic combiners. Overall, there are however two major differences between the results observed for Connected Components and PageRank. First, the best sequential performance is now achieved by both *GraphChi* and *Ligra*. Second, the thread scalability of *iPregel* is as good as *Ligra*'s on all graphs.

Finally, the timings collected in SSSP, presented in the right column of Fig. 11, show patterns found in the timings gathered for the Connected Components. Indeed, although *Ligra* remains several times faster than *iPregel*, the performance difference diminishes as the size of the graph increases. In-between stands GraphChi, faster than *iPregel* on low numbers of threads but due to its poor scalability it eventually falls behind as the number of threads increases. Also, we observe that *iPregel* continues to exhibit a thread scalability as good as that of *Ligra*.

---

[13] The capacity to provide performance gains when the number of threads increases.
[14] The capacity to provide performance gains when the size of the graph increases.

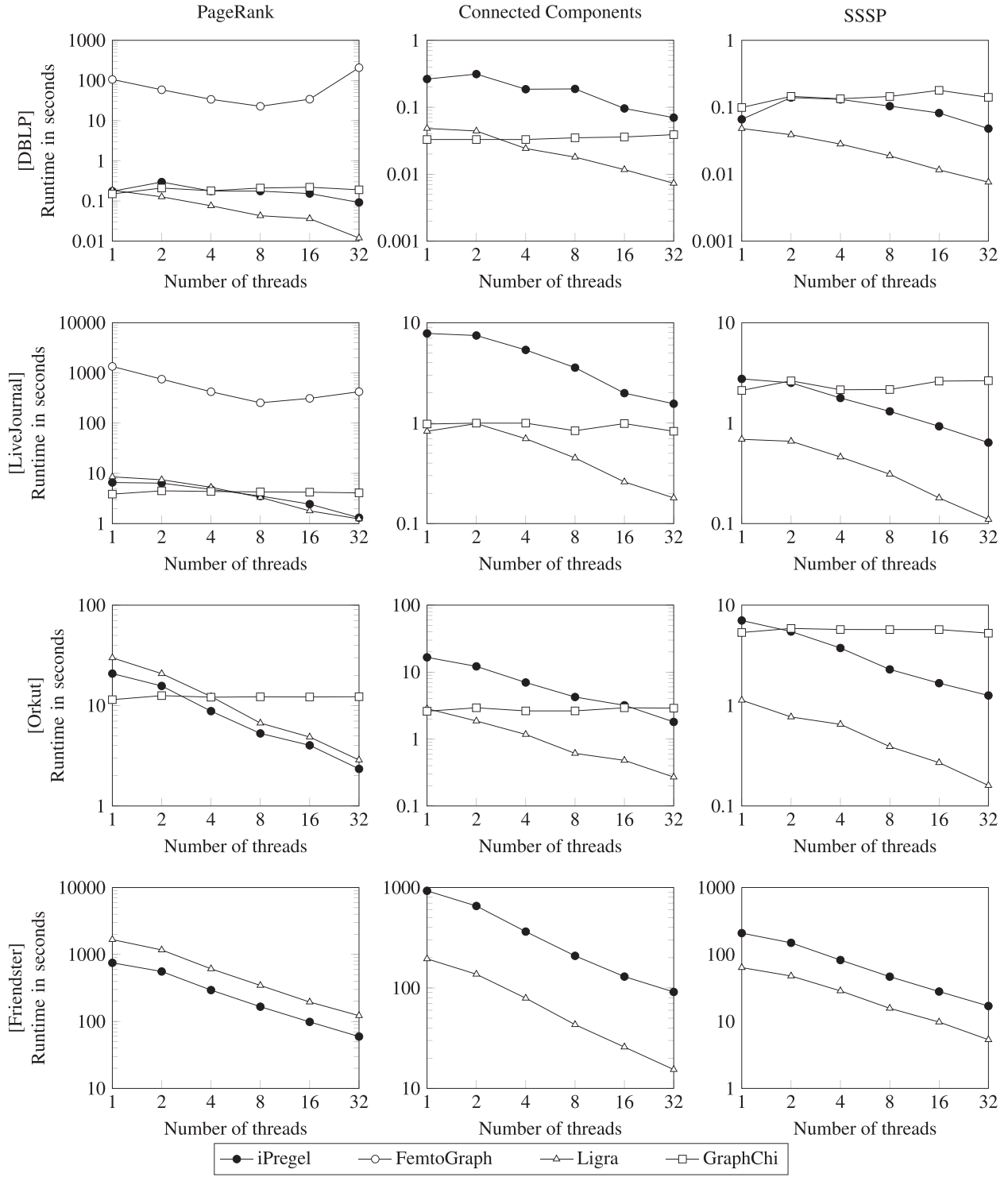

**Fig. 11.** Evolution of iPregel, Ligra, GraphChi and FemtoGraph runtimes against the number of nodes used, for each benchmark application, per graph.

### 8.2. Memory footprint

The memory footprints collected are reported in Table 3. We observe that *FemtoGraph* is up to 100 times less efficient than *iPregel*, eventually resulting in an out-of-memory failure for Friendster. The high memory overhead generated by *FemtoGraph* is partly due to the lack of message combination. Indeed, each vertex is provided with a mailbox that contains space for 100 messages while *iPregel* mailboxes only store the combined message, as introduced in Section 4.3.3. When processing the 65 million vertices of the Friendster graph (see Table 1), the *FemtoGraph* mailbox design requires 26GB[15] while that of *iPregel* uses 0.26GB. In addition to causing message losses when a vertex receives more than 100 messages, the *FemtoGraph* design results in wasted memory for vertices that receive fewer than 100 messages. Finally, we report that we were not able to process Orkut with *FemtoGraph* due to an abnormal termination that the debugging information provided by *FemtoGraph* does not permit us to explain.

---

[15] 65 million vertices storing 100 4-byte integers each.

**Table 3**
Maximum resident set size of each framework tested across all graphs processed, for each application executed, in Gigabytes. (Abbreviations used: CC = Connected Components, ABT = Abnormal Termination, OOM = Out Of Memory, FDO = File Descriptor Overflow).

|  | Graph | iPregel | FemtoGraph | GraphChi | Ligra |
|---|---|---|---|---|---|
| PageRank | DBLP | 0.07 | 3.26 | 0.07 | 0.04 |
|  | Live Journal | 0.48 | 51.95 | 1.41 | 0.51 |
|  | Orkut | 1.08 | ABT | 3.91 | 1.10 |
|  | Friendster | 20.45 | OOM | FDO | 21.43 |
| CC | DBLP | 0.15 | – | 1.06 | 0.03 |
|  | Live Journal | 0.42 | – | 2.49 | 0.48 |
|  | Orkut | 1.03 | – | 7.58 | 1.07 |
|  | Friendster | 20.94 | – | FDO | 20.45 |
| SSSP | DBLP | 0.14 | – | 0.10 | 0.02 |
|  | Live Journal | 0.47 | – | 2.49 | 0.42 |
|  | Orkut | 1.07 | – | 7.57 | 1.04 |
|  | Friendster | 19.91 | – | FDO | 18.19 |

According to Table 3, *GraphChi* is approximately 40 times more memory efficient than *FemtoGraph*. It produces a memory footprint that is within the same order of magnitude than *iPregel*. Nonetheless, *GraphChi* remains between 3 and 6 times less memory efficient on average. Despite providing an in-memory version of several applications, *GraphChi* remains a framework tailored for out-of-core computations, and it is therefore understandable that its memory usage is not as optimised as that of a pure in-memory framework like *iPregel* or *Ligra*.

Finally, we observe in Table 3 that the memory footprint of *Ligra* is similar to that of *iPregel*. In the majority of experiments, the difference is smaller than 60MB. The maximum difference in favour of *Ligra* is for SSSP on Friendster, where its memory footprint is 1.72GB (or 9%) smaller than that of *iPregel*. Conversely, running PageRank on the Friendster graph is where *iPregel* makes the biggest difference in its favour with 20.45GB against 21.43GB for *Ligra*; saving 0.98GB (approximately 5%). Among the two frameworks, the best in terms of memory efficiency depends entirely on the benchmark and graph being processed, *Ligra* proves to be more efficient than *iPregel* 7 times, while the contrary is observed 5 times. As a consequence, *iPregel* manages to provides a vertex-centric interface with a memory footprint as competitive as its non-vertex-centric counterpart.

*8.3. Programmability*

In this section, we evaluate the programmability of frameworks tested by comparison against the vertex-centric interface exposed in Pregel. Although Pregel is available within Google exclusively, its implementations for multiple benchmarks used in this paper are given in the original paper [1].

The PageRank application is the only one implemented by all four frameworks considered in this paper, so it was selected as the reference application. The PageRank implementation using the original Pregel framework is illustrated in Fig. 12, taken from [1]. We observe 3 characteristics that we use as evaluation criteria:

1. A vertex-centric interface; representing the fundamental advantage of the Pregel API with regard to programmability.
2. Encapsulated vertex data, that is, data specific to vertices are stored in vertices themselves, such as the rank for PageRank. This contrasts with another possible approach where vertices would fetch their rank from a global structure shared across all vertices. The latter however requires the user to be aware of the underlying addressing algorithm between a vertex identifier and the corresponding position in the global structure. As a result, encapsulating vertex attributes improves programmability by letting the framework handle the vertex addressing while exposing a less error prone programming interface to the user.
3. The completion of a vertex is expressed via a halting function. This is the cornerstone of vertex selection and algorithm termination, yet it requires very little work from the user: calling the halting function.

```
void Compute(MessageIterator* msgs) {
  if(superstep() >= 1) {
    double sum = 0;
    for(; !msgs->Done(); msgs->Next())
      sum += msgs->Value();
    *MutableValue() = 0.15 / NumVertices()
                    + 0.85 * sum;
  }
  if(superstep() < 10) {
    const int64 n = GetOutEdgeIterator().size();
    SendMessageToAllNeighbors(GetValue() / n);
  } else {
    VoteToHalt();
  }
}
```

**Fig. 12.** Compute function for PageRank in Pregel.

**Table 4**
Evaluation of frameworks considered against the programmability criteria defined from the Pregel implementation of PageRank. (Abbreviations used: IP = iPregel, FG = FemtoGraph, GC = GraphChi, LI = Ligra).

| Framework | IP | FT | GC | LI |
|---|---|---|---|---|
| Vertex-centric interface | Yes | Yes | Yes | No |
| Encapsulated attributes | Yes | Yes | No | No |
| Vertex halting | Yes | Yes | No | No |

```
void compute(queue<message*, fixed_sized<true>>*
    messages) {
  if(graph->superstepcount >= 1) {
    double sum = 0;
    message* m;
    while(messages->pop(m))
      sum += m->data;
    data->weight = 0.15 / graph->size()
                 + 0.85 * sum;
  }
  if(graph->superstepcount < 10) {
    const long n = outEdges.size();
    sendMessageToNodes(neighbors,
                       data->weight / n);
  }
  else voteToHalt();
}
```

**Fig. 13.** Compute function for PageRank in FemtoGraph.

In Table 4, we observe that *iPregel* and *FemtoGraph* fulfil all three programmability criteria. As we can see in their implementations given in Figs. 8 and 13, they clearly offer a highly abstracted vertex-centric interface, vertex-specific information are encapsulated in the vertices and the halting mechanism is invoked by vertices using a simple function call.

*GraphChi* too provides a vertex-centric interface. However, we observe in its implementation of PageRank in Fig. 14 that vertex ranks are contained in a single array *pr*. As a result, the user is in charge of handling the vertex addressing, and they have to manipulate this global structure from a centralised view and not vertex-centric. In addition, *GraphChi* performs the vertex selection via a

```
void update(graphchi_vertex<VertexDataType,
  EdgeDataType>& v, graphchi_context& ginfo) {
  if(ginfo.iteration == 0)
      pr[v.id()] = 1.0 / ginfo.nvertices;
  else if(ginfo.iteration > 0) {
    float sum = 0.0;
    for(int i = 0; i < v.num_inedges(); i++)
      sum += pr[v.inedge(i)->vertexid];
    pr[v.id()] = 0.15 / ginfo.nvertices
               + 0.85 * sum;
    if(v.outc > 0)
      pr[v.id()] /= v.outc;
  }
  if(ginfo.iteration < 10)
    v.set_data(v.outc > 0
               ? pr[v.id()] * v.outc
               : pr[v.id()]);
}
```

**Fig. 14.** Compute function for PageRank in GraphChi.

```
template <class vertex>
void Compute(graph<vertex>& GA, commandLine P) {
  long maxIters = 10, iter = 0;
  const intE n = GA.n;
  double one_over_n = 1/(double)n;
  double* p_curr = newA(double,n);
  {parallel_for(long i=0;i<n;i++)
    p_curr[i] = one_over_n;}
  double* p_next = newA(double,n);
  {parallel_for(long i=0;i<n;i++)
    p_next[i] = 0;}
  bool* frontier = newA(bool,n);
  {parallel_for(long i=0;i<n;i++)
    frontier[i] = 1;}
  vertexSubset Frontier(n,n,frontier);
  while(iter++ < maxIters) {
    edgeMap(GA,Frontier,PR_F<vertex>(
      p_curr,p_next,GA.V),0, no_output);
    vertexMap(Frontier,PR_Vertex_F(
      p_curr,p_next,0.85,n));
    vertexMap(Frontier,PR_Vertex_Reset(p_curr));
    swap(p_curr,p_next);
  }
  Frontier.del(); free(p_curr); free(p_next);
}
```

**Fig. 15.** Compute function for PageRank in Ligra.

```
template <class vertex> struct PR_F {
  double* p_curr;
  double* p_next;
  vertex* V;
  PR_F(double* _p_curr, double* _p_next,
    vertex* _V)
    : p_curr(_p_curr), p_next(_p_next), V(_V) {
  }

  inline bool update(uintE s, uintE d) {
    p_next[d] += p_curr[s]
            / V[s].getOutDegree();
    return 1;
  }

  inline bool updateAtomic(uintE s, uintE d) {
    writeAdd(&p_next[d],
            p_curr[s]/V[s].getOutDegree());
    return 1;
  }

  inline bool cond(intT d) {
    return cond_true(d);
  }
};

struct PR_Vertex_F {
  double damping;
  double addedConstant;
  double* p_curr;
  double* p_next;
  PR_Vertex_F(double* _p_curr, double* _p_next,
            double _damping, intE n)
    : p_curr(_p_curr), p_next(_p_next),
      damping(_damping),
      addedConstant((1 - _damping)
                *(1 / (double)n))){
  }

  inline bool operator () (uintE i) {
    p_next[i] = damping * p_next[i]
            + addedConstant;
    return 1;
  }
};

struct PR_Vertex_Reset {
  double* p_curr;
  PR_Vertex_Reset(double* _p_curr)
    : p_curr(_p_curr) {
  }

  inline bool operator () (uintE i) {
    p_curr[i] = 0.0;
    return 1;
  }
};
```

**Fig. 16.** Additional user-defined structures needed by the PageRank compute function in Ligra.

vertex scheduler, which can be disabled for algorithms such as PageRank, resulting in no halting mechanism available at vertex-level. Although the algorithm termination is "based on all vertices voting to halt" according to Pregel [1], in the *GraphChi* version of PageRank it is determined in the main function, where a maximum number of iterations is defined. PageRank has a particularity; all vertices run at every superstep. However, this is not the case for most algorithms, including the Connected Components and SSSP, therefore requiring a vertex selection mechanism. In *GraphChi*, this is achieved via a vertex scheduler that must be explicitly enabled or disabled by the user, then called in user code when processing each vertex. Indeed, for an algorithm that requires vertex selection like SSSP, vertices that send a message must then explicitly call the scheduler and schedule the recipient vertex for execution. This approach exposes implementation-level details to the user. By contrast, *iPregel* abstracts the vertex selection behind the call to the halting function. In addition, the selection bypass optimisation presented in Section 4.3.1 is enabled via a compilation flag, without requiring a modification in the user application source code.

That allows the user to rely on a consistent programming interface across all applications, unlike *GraphChi* where, for instance, vertices do not halt in PageRank whereas they do in SSSP, and sending a message must be followed by an explicit schedule of the recipient vertex in SSSP, while it does not in PageRank.

Finally, Fig. 15 illustrates the *Ligra* implementation of PageRank. For fairness, we removed the source code section that was calculating the convergence of PageRank results since other frame-works (including *iPregel*) do not do this. This deletion is also beneficial to *Ligra* from a programmability perspective as it reduces the amount of code written and hides the details about convergence calculations from the code. Nonetheless, we observe none of the criteria presented in Table 4, although this is understandable for a framework that is not vertex-centric but vertex-subset-centric. The source code provided explicitly exposes parallelism to the user in two aspects. First, syntactically, as we can see with the use of parallel for loops wrapped in curly brackets. Second, semantically, as *Ligra* states in [5], the function provided to edgeMap "*can run in parallel, so the user must ensure parallel correctness*". In other words, the user is in partially responsible for the thread-safety of *Ligra*. In addition, the iterative structure of the computation as well as dynamic memory allocations and deallocations are done directly by the user, as shown in Fig. 15. This a price the designers of *Ligra* have accepted in order to obtain more performance, but we argue that such concerns are too low-level for the user. Furthermore, we observe that the *compute* function only outlines the general computation flow. The edge map and vertex map functions must be defined by the user as well, which are given in Fig. 16 for PageRank. As we can see, the total amount of code that must be written by the user greatly exceeds that for *iPregel*. As explained in Section 4.3.2, *Ligra* provides atomic combination as an additional optimisation, which is enabled by writing a second version of the update function. This requires the user to be aware of the atomicity potential of its combination operation, as well as being able to implement it atomically using the *Ligra* functions provided.

## 9. Conclusions and future work

Our initial observation that programmability suffers from optimisations made for memory efficiency and performance is illustrated in the results collected. Preserving the vertex-centric programmability leads *FemtoGraph* to be up to 17,000 times and 700 times slower than *Ligra* and *GraphChi* respectively, in addition to resulting in a memory footprint up to orders of magnitude bigger.

Experiments demonstrate that our framework, *iPregel*, provides the best of both worlds. By developing optimisations that do not hinder programmability, *iPregel* has been able to bridge the weaknesses of *FemtoGraph* with regard to memory efficiency and performance, without sacrificing programmability. This statement can not be said about any other vertex-centric framework including *FemtoGraph*, *GraphChi* and *Ligra*. The memory efficiency of *iPregel* equals that of *Ligra* which was the most memory efficient framework tested. *iPregel* is also up to 100 times more memory efficient than *FemtoGraph*, and up to 7 times more memory efficient than *GraphChi*. This additional memory efficiency allows *iPregel* to process graphs that *FemtoGraph* cannot because its memory footprint exceeds the available memory. Regarding performance, the maximum speedup of *GraphChi* or *Ligra* over *iPregel* is at most 10, which is up to 1700 times less than the speed-up they can achieve over *FemtoGraph*. The performance observed on *iPregel* is up to 2300 times better than that of *FemtoGraph*. At worst, *iPregel* remains 70 times faster than *FemtoGraph*. On PageRank, *iPregel* even manages to provide the best performance overall, outperforming its programmability-hindered counterparts as well.

The multifaceted analysis of this paper demonstrates that our framework *iPregel* overcomes the fundamental compromise in vertex-centric frameworks by reaching a point where vertex-centric programmability no longer impacts memory efficiency, and can result in a limited performance loss, no loss at all or even a performance gain. This makes *iPregel* the first shared memory vertex-centric framework able to scale to a multi-billion edge graph without sacrificing vertex-centric programmability.

Further improvements of *iPregel* could include the design and implementation of atomic combiners that do not hinder the programmability exposed to the user. Also, the multi-threaded performance observed would certainly benefit from additional investigations in load-balancing strategies and work stealing techniques. Finally, porting *iPregel* to a distributed memory architecture is a third potential direction, which may lead future efforts.

## Acknowledgements

We thank the reviewers for their helpful feedback and suggestions. This research was supported by the International Internship Program of the National Institute of Informatics of Tokyo and the UK Engineering and Physical Sciences Research Council [grant number EP/L01503X/1, CDT in Pervasive Parallelism].

## Supplementary material

Supplementary material associated with this article can be found, in the online version, at doi:10.1016/j.parco.2019.04.005.